\documentstyle[prb,preprint,aps]{revtex}
\begin{document}
\title{Spin-Wave-Spin-Wave Interaction and the Thermodynamics of the
Heisenberg Spin Chain}

\author{Michael Teitelman\thanks{email: teitel$@$appl.nnov.su}}
\address{ Department of Solid State Physics  \\
          Institute for Applied Physics,
          Russian Academy of Sciences \\
          603600, Nizhny Novgorod, 46 Uljanov St., RUSSIA}
\date{\today}

\maketitle
\begin{abstract}
The low-temperature free energy of the spin S quantum Heisenberg
ferromagnetic chain in a strong magnetic field is obtained in a
two-particle approximation by using exact solution of two-spin-wave
problem. The result is beyond the perturbation theory because it
incorporates the both bound and scattering state contributions, and the
scattering effect is essential as well as the bound state one. In
particular the main temperature renormalization of an exchange constant
is found to be linear in temperature instead $T^{3/2}$ corresponded to
the perturbation theory result.

\end{abstract}
\pacs{PACS: 75.10.Jm, 75.30.Ds, 05.30.-d, 67.40.Db}
\narrowtext

The low-dimensional spin systems show many special features in the
thermodynamics. Some of them are the result of a large number of
excitations there, for example the destruction of a long-range order
that well-known Mermin and Wagner theorem states \cite{Merm-Wag}. Also
these features are often coupled with the failure of the free-spin-wave
based consideration for the low-dimensional systems. In this
communication we would like to note that the exchange interaction of
even two spin waves in Heisenberg chain shows an effect which is beyond
perturbation theory. In fact this property is connected with well-known
peculiarity of low-dimensional quantum mechanical problem such as the
bound state formation and the strong scattering at a long-wave limit in
a weak potential \cite{Landau}. Though as spin $S\rightarrow\infty$ it
is a weak coupling limit the perturbation theory is not correct for the
long-wave excitations which dominate at the low temperatures. The
consideration of the $1/S$ corrections was noted early to be essential
in low-dimensional spin systems \cite{Chub,Mills}.

We explore pure exchange Heisenberg ferromagnetic spin chain with
arbitrary spin $S$ and a strong magnetic field in a limit of low
temperatures. The Hamiltonian is the following

\begin{equation}
H=-J\sum_{<i,j>}{\bf S}_i {\bf S}_j -h\sum_i S^z_i .
\label{H}
\end{equation}
In the first term we sum up the different pair of the nearest neighbor
sites where quantum spins ${\bf S}_i$ are located.
When $S=1/2$ this system is solved by Bethe Ansatz method
{}~\cite{Bethe}. As $S$ is larger there is not an exact solution. On the
other hand it is supposed the perturbation theory is correct as
$S\rightarrow\infty$ \cite{Holstein}. We show below this statement is
wrong for a spin chain. Other known methods also do not allow consider
specific features of low-dimensional systems \cite{Takahashi,Auerbach}.

The energy of a ferromagnetic spin system caused an applied magnetic
field depends on a number of excitations only. That allows us to
consider separately the states with one, two spin deviations and so on.
We are interesting the correction to the free energy due to interaction
which arises from two-spin-wave sector at first.  The two-spin-wave
problem is exactly solved for arbitrary spin $S$ \cite{Wortis}. The
energy spectrum of two spin waves in the finite chain of size $N$
follows from the eigenvalue equation\cite{Wortis}

\begin{equation}
D(p,q)=1 - \frac{1}{N}\sum_{k}\frac{2J\cos(k)(\cos(k)-\cos(p/2))}
{2S\cos(p/2)(\cos(k)-\cos(q))} = 0,
\label{D}
\end{equation}
where $k=2\pi n/N$, $n=0, 1, .., N-1$, and
$p$ is center mass momentum of two-spin-wave state. The energy
spectrum of the state is parametrized by $q$ as follows

\begin{equation}
E(p,q)=4SJ(1-\cos(p/2)\cos(q)) + 2h.
\end{equation}

Let us consider function $D$ vs $q$ at the fixed center mass momentum $p$.
While the chain is finite $D(q)$ is the meromorphic function
with $N$ poles and $N$ zeros, where the poles correspond to free spin
waves and the zeros correspond to real energy levels. As
$N\rightarrow\infty$ function $D$ changes its analytical properties. In
the limit of a large $N$ it gives the isolated zero beyond continuous
spectrum and a cut presented scattering state. In this limit the
difference between scattering state energy levels and the
free-spin-wave energy levels is $\sim 1/N$. Just this small value gives
a perturbation of the free energy by spin-wave-spin-wave scattering,
because the number of scattering states is $\sim N^2$ and therefore the
scattering contributes in the order $N$.

The finite sum in (\ref{D}) is calculated directly and we obtain

\begin{equation}
D(p,q)=1 - \frac{1}{2S}\left(1-\frac{\cos(q)}{\cos(p/2)}\right)
\left(1+\cot(q)\cot(\frac{Nq}{2})\right) .
\label{Dq}
\end{equation}

The state out of continuum corresponds imaginary parameter $q$. We use
$q=i\alpha$. Then as $N\rightarrow\infty$ the Eq.(\ref{Dq}) gives
the bound state condition which is the same as in the ref.~
\cite{Wortis}.

\begin{equation}
1 - \frac{1}{2S}\left(1-\frac{\cosh(\alpha)}{\cos(p/2)}\right)
\left(1-\coth(\alpha)\right)=0 .
\label{Da}
\end{equation}

In the long-wave limit we obtain $\alpha=(p/2)^2/4S+O(p^4)$. Thus the
coupling energy is $E_b(p)-E(p,0)=-4SJ(p/2)^4/32S^2$.

To find the perturbation spectrum of the continuum states we use
$q=2\pi (n+\Delta)/N$ where $n$ is integer and $\Delta$ characterizes
the energy level shift. Then from the Eq.(\ref{Dq}) we obtain the
equation for $\Delta$

\begin{equation}
1 - \frac{1}{2S}\left(1-\frac{\cos(q)}{\cos(p/2)}\right)
\Big(1+\cot(q)\cot(\pi\Delta(q))\Big)=0 .
\label{Del}
\end{equation}
Therefore
\begin{equation}
\Delta(q)=\frac{1}{2}-\frac{1}{\pi}\arctan\left\{
\tan(q)\left(\frac{2S}{1-\frac{\cos(q)}{\cos(p/2)}}-1\right)
\right\}
\label{Del(q)}
\end{equation}

The function $\Delta(q)$ is essentially different in the case when
$S=1/2$ or $S>1/2$. This function is periodic for $S>1/2$ and
multivalued for the half spin. In the last case
$\Delta(\pi)-\Delta(0)=1$, that provides correct reduction of the
number of states for $S=1/2$.

After elementary transformations by keeping the terms order $N$ we
obtain the spin-chain free energy $f$ which forms from one- and
two-spin-wave states by the temperature $T$

$$
-\frac{f}{T}=N e^{-h/T} \int\limits_{-\pi}^{\pi}\frac{dk}{2\pi}
e^{-\varepsilon_1(k)/T}
-\frac{N}{2}e^{-2h/T} \int\limits^{\pi}_{-\pi}\frac{dp}{2\pi}
\int\limits^{\pi}_{0}dq \frac{d\Delta}{dq}e^{-\varepsilon_2(p,q)/T}+
$$
\begin{equation}
\frac{N}{2}e^{-2h/T} \int\limits^{\pi}_{-\pi}\frac{dp}{2\pi}
e^{-\varepsilon_b(p)/T} ,
\label{f}
\end{equation}
where $\varepsilon_1(k)=2SJ(1-\cos(k))$ and
$\varepsilon_2(p,q)=4SJ(1-\cos(p/2)\cos(q))$ are one- and two-spin-wave
energy and $\varepsilon_b(p)$ is the bound state energy without the
magnetic field term. The factor $1/2$ in the two-spin-wave terms follows
from the permutation symmetry of this state. We note zero $q$ state is
not realized as follows from the Eq.(\ref{Del}).

We calculate the scattering term using the Eq.(\ref{Del(q)}) and
bound state term using the bound state energy from the Eq.(\ref{Da}).
Then we obtain

\begin{eqnarray}
-\frac{f_{scatt}}{NT}=e^{-\frac{2h}{T}}\left\{
-\frac{1}{2\sqrt{2}}\left(\frac{T}{4\pi
SJ}\right)^{1/2}\left[1+\frac{2\pi}{16}\left(\frac{T}{4\pi
SJ}\right)\left(1+\frac{3}{(2S)^2}\right)+\right. \right.
\nonumber\\
\left. \left.
\frac{2\pi\sqrt{2}}{16(2S)}
\left(\frac{T}{4\pi
SJ}\right)^{3/2}\left(1+\frac{4}{2S}-\frac{15}{2(2S)^2} \right)
\right]+O(T^{5/2})
\right\} ;
\label{scatt}
\end{eqnarray}
\begin{equation}
-\frac{f_{bound}}{NT}=e^{-\frac{2h}{T}}\left\{
\frac{1}{\sqrt{2}}\left(\frac{T}{4\pi
SJ}\right)^{1/2} + \frac{2\pi}{16\sqrt{2}}\left(\frac{T}{4\pi
SJ}\right)^{3/2}\left(1+\frac{3}{(2S)^2}\right)+O(T^{5/2})
\right\}.
\label{bound}
\end{equation}
We note that the bound and scattering contributions are the same order.
The free-spin-wave theory gives
\begin{equation}
-\frac{f_{FSW}}{NT}=-\int\limits^{\pi}_{-\pi}\frac{dk}{2\pi}\ln
\left(1-e^{-\varepsilon_1(k)/T}\right)=\int\limits^{\pi}_{-\pi}\frac{dk}{2\pi}
\left(e^{-h/T-\varepsilon_1(k)/T}+\frac{1}{2}e^{-2h/T-2\varepsilon_1(k)/T}
+...\right).
\label{fsw}
\end{equation}
Thus the deviation from free-spin-wave free energy is
\begin{eqnarray}
-\frac{f-f_{FSW}}{N}=T\left\{\frac{2\pi}{32\sqrt{2}}\frac{3}{(2S)^2}
\left(\frac{T}{4\pi SJ}\right)^{3/2}-\frac{2\pi}{32(2S)}
\left(\frac{T}{4\pi SJ}\right)^{2}
\left(1+\right. \right. \nonumber\\
\left. \left.
\frac{4}{2S}-\frac{15}{2(2S)^2}\right)\right\}
e^{-\frac{2h}{T}}+
O(T^{7/2})
e^{-\frac{2h}{T}} .
\label{Df}
\end{eqnarray}

We note this temperature expansion of the free energy does not contain
the $T^2$ term. It is in agreement with the results of the exactly
solved Heisenberg system for spin $1/2$\ \cite{Johnson}.
The term $T^3$ which corresponds to the perturbation theory is also
presented. It is interesting to note how this term depends on spin
value $S$. When $S=1/2$ it has the same sign as a main contribution and
for all others spins it has opposite sign.

It is interesting to consider the temperature
renormalization of an exchange parameter which characterizes the
temperature correction to the energy spectrum. As defined this
renormalization reduces the real free energy to the free-spin-wave
form. So we obtain

\begin{equation}
\frac{\Delta J}{J}=-\frac{3T}{16\sqrt{2}J(2S)^3}e^{-\frac{h}{T}}
\left(1+\frac{2h}{3T+2h}\right) .
\end{equation}

Thus the spin-wave-spin-wave exchange interaction gives a linear
temperature factor in the effective exchange parameter. A similar
result follows from the consideration of a classical spin system. On
the other hand since the leading temperature term contains extra
factor $1/S$ then in a large spin limit at the fixed temperature
it will be suppressed and other contributions will become more
essential.  This fact means that low-temperature expansion is divergent
when spin is large. It is easy to see the using of the $1/S$ expansion
for the exact scattering spectrum in the Eq.(\ref{Del(q)}) gives the
divergences in the expression for the free energy.
It explains the failure of the perturbation theory for a spin
chain.


\begin{thebibliography}{99}
\bibitem{Merm-Wag}N.D.Mermin and H.Wagner, Phys.Rev.Lett.{\bf 17},
1133(1966).

\bibitem{Landau}L.D.Landau and E.M.Lifshitz, "Quantum Mechanics";
Pergamon Press.

\bibitem{Chub}Y.A.Kosevich and A.V.Chubukov, Zh.Eksp.Teor.Fiz.{\bf 91},
1105(1986)
[Sov.\ Phys.\ JETP {\bf 64}, 654 (1986)]
; D.~V.~Khveshchenko and A.~V.~Chubukov,
Zh.\ Eksp.\ Teor.\ Fiz.\ {\bf 93}, 1904 (1987)
[Sov.\ Phys.\ JETP {\bf 66}, 1088 (1987)]
{}.

\bibitem{Mills}R.P.Erickson and D.L.Mills, Phys.Rev.B {\bf 46}, 861
(1992);
R.W.Wang and D.L.Mills, Phys.\ Rev.\ B {\bf 48}, 3792
(1993-I).

\bibitem{Bethe}H.A.Bethe, Z. Phys. {\bf 71}, 205 (1931).

\bibitem{Holstein}T.Holstein and H.Primakoff, Phys.Rev. ser.2 {\bf 58},
1098 (1940).

\bibitem{Takahashi}M.Takahashi, Prog.Teor.Phys.Suppl. {\bf 87},
233(1986).

\bibitem{Auerbach}D.P.Arovas and A.~Auerbach, Phys. Rev. B {\bf 38},
316 (1988); A.~Auerbach and D.P.~Arovas, J. Appl. Phys. {\bf 67},
5734 (1990).

\bibitem{Wortis}M.Wortis, Phys. Rev. {\bf 132}, 85 (1963).

\bibitem{Johnson}J.D.Johnson and B.M.McCoy, Phys.Rev. A {\bf 6},
1613(1972).

\end{thebibliography}
\end{document}